# Feature Importance in the Context of Traditional and Just-In-Time Software Defect Prediction Models

Susmita Haldar
*School of Information Technology*
*Fanshawe College*
London, Canada
shaldar@fanshawec.ca

Luiz Fernando Capretz
*Department of Electrical and Computer Engineering*
*Western University*
London, Canada
lcapretz@uwo.ca

*Abstract*—Software defect prediction models can assist software testing initiatives by prioritizing testing error-prone modules. In recent years, in addition to the traditional defect prediction model approach of predicting defects from class, modules, etc., Just-In-Time defect prediction research, which focuses on the change history of software products is getting prominent. For building these defect prediction models, it is important to understand which features are primary contributors to these classifiers. This study considered developing defect prediction models incorporating the traditional and the Just-In-Time approaches from the publicly available dataset of the Apache Camel project. A multi-layer deep learning algorithm was applied to these datasets in comparison with machine learning algorithms. The deep learning algorithm achieved accuracies of 80% and 86%, with the area under receiving operator curve (AUC) scores of 66% and 78% for traditional and Just-In-Time defect prediction, respectively. Finally, the feature importance of these models was identified using a model-specific integrated gradient method and a model-agnostic Shapley Additive Explanation (SHAP) technique.

## I. INTRODUCTION

Software defect prediction (SDP) models attempt to predict bugs early in the life cycle of software projects and assist test managers in deciding which artifacts to focus on for testing. This can lead to efficient test planning, improving software development practices [1], [2], and improved software quality. In earlier research, the primary goal of the SDP model was to identify individual error-prone modules, files, or programs. This is referred to as the traditional approach in this study. In the current trend of defect prediction, Just-In-Time (JIT) defect predictions are [3], [4] being explored which identifies project-level commits that are likely to induce bugs.

The JIT models are trained based on a set of previous change revisions and whether that revision introduced a bug to the software referred to as bug-inducing commits. Such commit-level predictions help project leads prioritize their testing of the riskiest commits [5]. On the other hand, traditional defect prediction studies, predict the modules or programs that are likely to be error-prone based on certain metrics available in the source code such as total lines of code, number of functions, complexity in the code, etc. From a test manager's perspective, both types of SDP models can bring value in test planning based on which phase of the software life cycle they are in.

It is crucial to understand which features contribute to these SDP models to explain the model outcomes and to provide the required inputs for training these models. Existing research studies have looked at either the traditional approach for finding defect-prone modules or the Just-In-Time approach to identify bug-inducing commits. However, a combined approach for identifying features in both traditional and JIT SDP models has not been explored yet. Such an integrated approach would assist test managers with early test planning in the project's life cycle as well as before production releases.

Several machine learning and deep learning algorithms have been effectively applied in various software defect prediction studies, and the utilization of deep learning algorithms has shown promise [4], [6] in identifying defect-prone artifacts. It needs to be explored if the same algorithms are equally effective in both of these SDP approaches to promote the generalization of these algorithms with different types of datasets. Often deep learning algorithms are considered as black-box models whereas the model itself does not contain the explanation of the predictions. Additional techniques are usually applied to understand the reasoning behind predictions.

To assist the research community in furthering the defect prediction studies, the following research questions will be explored in this study:

RQ1 How do traditional defect prediction models compare to just-in-time defect prediction models in terms of prediction accuracy and reliability when using deep learning algorithms?

RQ2 Which features seem important when developing these defect prediction models? Is there any feature that intersects between these SDP models?

The rest of the paper has been organized as follows. Section II does a literature survey on the traditional and JIT defect prediction models. Section III describes the methodology used for implementing these defect predictions. Section IV presents the results of the experiment followed by a discussion of the findings and limitations of this study. Finally, section V concludes this study with closing remarks and the future work of this study.

## II. BACKGROUND AND LITERATURE SURVEY

Traditional SDP models often use static code metrics that are directly extracted from source code [7]. For Just-in-time defect prediction, collecting process metrics is common [8].

Process metrics measure the change information of the source code during a period which can be extracted from the Source Code Management system based on historical information on changes in source code over time.

To build these SDP models, researchers have used various classification techniques such as Support Vector Machine (SVM) [9], Deep learning [6], Random Forest (RF) [10], Logistic Regression [8], etc. The researchers are currently focusing on how these models can be explained based on feature importance and other factors [10]. Our previous work focused on interpretability aspects of software defect prediction studies using ISBSG and Promise datasets [11], [12].

Bludau et al. [13] analyzed existing feature sets along with proposing two new features effective for JIT defect prediction studies. They recommended selecting features with care as the feature importance varies among projects. Our study will investigate the existing features found in the literature for both traditional and JIT defect prediction models to be able to compare their performance and effectiveness.

Rajbahadur et al. [14] investigated the feature importance ranking for model-specific and model-agnostic scenarios, and they found that the model-specific scenarios varied widely for feature ranking compared to model-agnostic approaches for traditional defect prediction models. In this study, we will apply both model-specific and model-agnostic approaches to identify feature importance.

In recent years, several open-source solutions have allowed researchers to reproduce the work in both traditional and JIT defect prediction areas. Grigoriou et al. [15], developed a large dataset by mining 148 projects from GitHub and sharing raw data in a public repository. Their work was on the traditional approach rather than applying JIT defect prediction. Apache JIT is another large dataset with 106,674 software changes applicable to defect prediction studies that have been shared in the public repository for Just-In-Time (JIT) defect prediction [16].

This study has utilized existing reusable datasets available in the literature to compare the model performance in each of these models and compare the important features that contributed to these models.

## III. METHODOLOGY

The methodology for this study has been outlined in Figure 1 which started with dataset selection followed by developing traditional and JIT defect prediction models using existing machine learning and deep Learning approaches. The performance of the deep learning algorithm of these models was compared. The feature importance was identified using a model-specific approach for neural networks and the same model was later interpreted with a model-agnostic technique referred to as SHAP.

Six different projects were selected from the Apache project. Three projects were used for the traditional defect prediction study available from the dataset shared by Grigoriou et al. [15] which was initially introduced by Jureckzo [17]. The names of the projects are Apache Camel version 1.6, Apache Tomcat, and Apache Ant version 1.7, and for JIT defect prediction studies the same Camel project along with Apache Kafka and Zookeeper from the ApacheJIT dataset was used [16].

Logistic Regression, Random Forest, and Deep Learning algorithms were applied to develop these models. The performance of Logistic Regression and Random Forest was compared to the performance of the deep learning algorithm to identify how the performance varies between traditional and JIT SDP models.

Next the findings from the Apache Camel project were considered for final evaluation as this project was common in both the JIT dataset and traditional defect prediction dataset.

As the apacheJIT dataset [16] has multiple projects, this work filtered out the selected projects during the implementation. Table I shows the features and descriptions selected for predicting bug-inducing comments. Variable 'Buggy' was the predicted variable, and the rest of the features were considered independent variables. During the feature pre-processing and cleaning step, the scikit-learn StandardScaler method [18], which standardizes features by removing the mean and scaling to unit variance, was applied. Less important features such as Commit ID, project name, author-date fields, etc. were removed.

TABLE I: Attribute description of JIT defect prediction dataset

| Features | Description |
|---|---|
| Fix | Whether or not the change is a defect fix |
| LA | Number of lines added |
| LD | Number of lines deleted |
| NF | Number of files touched, |
| ND | Number of directories touched |
| NS | Number of subsystems touched |
| Entropy | Distribution of modified code across each file |
| NDEV | Number of distinct developers touched files |
| Age | The average time from the last change |
| NUC | Number of unique changes in files |
| EXP | Number of change author experience |
| REXP | Developer's recent experience |
| SEXP | Developer's subsystem experience |
| Buggy | Selected commit is buggy or not |

Feature structures of the traditional defect prediction have been shown in Table II. The dependent variable 'bug' is a continuous variable that includes the total number of bugs per class. For the development of the traditional defect prediction model, the bug field was converted to a 0 when the class is not buggy and 1 otherwise.

Both of these approaches used binary classification for predicting defect-prone modules or commits. The selected machine learning, and deep learning algorithm assigned equal weights to both defective and non-defective classes to ensure representation from both categories was considered as the datasets were imbalanced. The developed deep learning model consisted of five fully connected dense layers. Dropout layers were added to prevent overfitting. The architecture had dense layers consisting of 64, 32, 20, and 10 neurons, each followed by a dropout layer. The final dense layer contained a single neuron for binary classification. Accuracy and AUC were used as evaluation metrics which are proven to work with binary classification problems for unbalanced datasets [19]. The project was implemented using Anaconda Jupyter Notebook utilizing Python scripts.

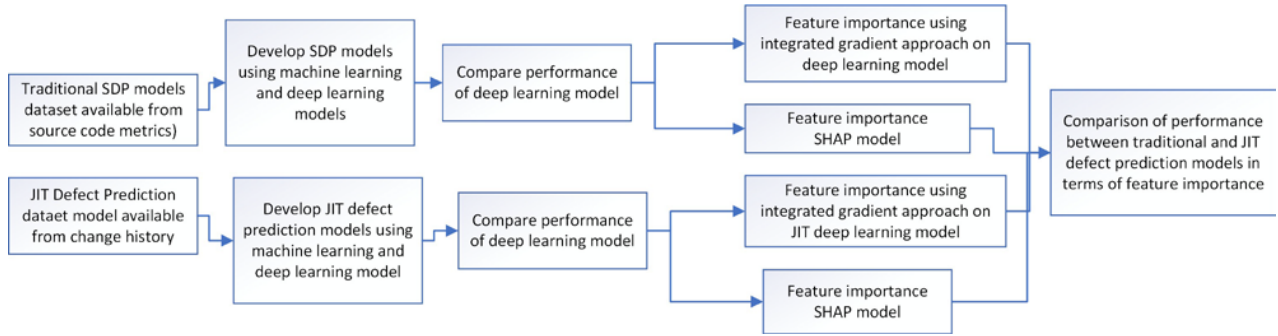

Fig. 1: Methodology for comparing the performance of traditional versus deep learning models

TABLE II: Features of SDP projects for predicting bugs of classes (traditional approach)

| Features | Description |
| --- | --- |
| cbo | Coupling between Objects |
| dit | Depth of Inheritance Tree |
| lcom | Lack of Cohesion of Methods |
| noc | Number of Children |
| rfc | Response For Class |
| wmc | Weighted Methods for Class |
| ca | Afferent couplings. |
| ce | Efferent couplings. |
| npm | Number of Public Methods |
| lcom3 | Lack Of Cohesion in Methods |
| loc | Lines Of Code |
| dam | Data Access Metrics. |
| moa | Measure Of Aggregation. |
| mfa | Measure of Functional Abstraction. |
| cam | Cohesion Among Methods of class. |
| ic | Inheritance Coupling. |
| cbm | Coupling Between Methods |
| amc | Average Method Complexity |
| max_cc | Maximum of the McCabe's cyclomatic complexity |
| avg_cc | Average of the McCabe's cyclomatic complexity |
| bug | Total number of bugs found in this class |

For measuring feature importance, the Integrated Gradient (IG) method [20] was used as it was not possible to measure the coefficient value of the independent features using a direct approach for the neural networks. IG is a method that helps us understand how much each feature contributes to a prediction made by a model, like a deep learning neural network. It is a feature attribution method for explaining predictions from a differential function F from the deep learning model [21].

Let's define a starting point called the reference point, $x'$, and a prediction made by the SDP model for this reference point, $F(x')$. For a new instance, $x$, IG calculates an attribution vector. This vector tells us how important each feature is for the prediction made for $x$. To get this attribution vector, IG looks at the path from the reference point $x'$ to the new observation $x$ in the input space. Along this path, it considers how the SDP model's prediction changes as each feature changes. It integrates these changes to figure out the overall contribution of each feature to the final prediction. The attribution vector would provide us with a simple breakdown of how much each feature weighs when the model predicts a specific input. The

Equation 1 computes an attribution vector for a new instance $x$ for a reference point $x'$.

$$IG_i(x) = (x_i - x'_i) \int_{a=0}^{1} \frac{\partial F(x' + a(x - x'))}{\partial x_i} da \quad (1)$$

Next, the feature importance was validated by SHAP [22], a model-agnostic technique that utilizes a game theory-based approach for identifying contributions. The feature importance scores from both of these approaches were normalized by dividing each score by the maximum score to bring them to a common unit.

## IV. RESULTS AND ANALYSIS

Table III shows the performance of the SDP models when applied in each of these projects. The column 'Number of Instances' shows the size of the dataset based on the number of records available in the project. It was observed that for the traditional SDP approach, the deep learning model obtained the highest accuracy for projects of Tomcat and Camel with scores of 80% and 90%. For the AUC score, however, the Random Forest algorithm performed better with values of 69% and 84% compared to 66% and 75% for these projects. Considering a balanced score from AUC and accuracy, the deep learning method performed well for larger datasets. Logistic regression did not outperform the performance of the other two classifiers. For the JIT defect prediction model, the performance of the Camel project which had the most number of instances, the deep learning algorithm performed best for accuracy with a score of 86%. In the Zookeeper project which has a relatively smaller dataset, the deep learning algorithm performed poorly compared to others. The random forest algorithm worked with the same consistency for all datasets. Our first research question wanted to explore the accuracy and reliability of deep learning algorithms for both traditional and JIT SDP models. In both instances, the deep learning algorithm outperformed in terms of accuracy for larger datasets with a slight reduction in AUC compared to the random forest model.

Our second research question was which features are important for these defect prediction studies, and to identify the intersection of the features between these two approaches. Figure 2 illustrates the order of feature importance for the traditional defect prediction models. The feature importance metrics show that lcom3 (lack of cohesion) contributes the most to the traditional SDP model. This is logical as it is desired to have high cohesion, and low coupling among application modules. The next attributes are avg_cc, cbm, ic (inheritance coupling), mfa, etc. The attribute avg_cc shows the average high cyclomatic complexity of the code is an important attribute for identifying error-prone modules. Software programs with high cyclomatic complexity are more complex and harder to understand, which can make it more difficult to identify and fix defects. On the other hand, the SHAP method showed a relatively different feature importance ranking for the same project as illustrated in Figure 3. The total line of code is shown as the top contributor to the model according to the SHAP technique followed by lcom, RFC, etc. In this circumstance, the feature ranking from SHAP can be more reliable as according to previous research [14], SHAP shows less variability than the model-specific approaches for feature rankings. For traditional SDP models, cbm (coupling between methods) is among the top 5 features according to both of the applied techniques.

TABLE III: Results obtained from running the selected algorithms

| Project Name | Number of instances | Logistic Regression ACC./AUC | Random Forest ACC./AUC | Deep Learning ACC./AUC | Applied SDP model approach |
|---|---|---|---|---|---|
| Camel 1.6 | 965 | 65/66 | 79/69 | 80/66 | Traditional |
| Tomcat | 848 | 74/81 | 89/84 | 90/75 | Traditional |
| Ant 1.7 | 748 | 78/76 | 82/82 | 80/82 | Traditional |
| Camel | 22700 | 72/77 | 81/83 | 86/78 | JIT |
| Kafka | 2384 | 81/88 | 85/91 | 82/88 | JIT |
| Zookeeper | 839 | 81/86 | 83/88 | 63/51 | JIT |

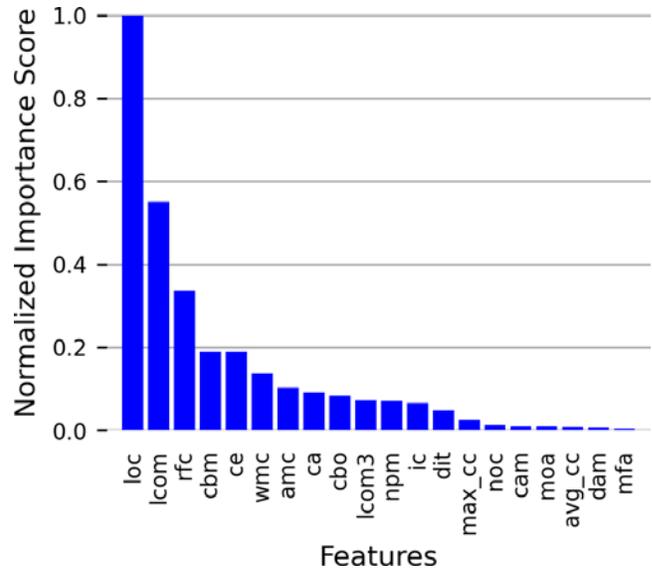

Fig. 3: Feature importance of the SDP model using SHAP.

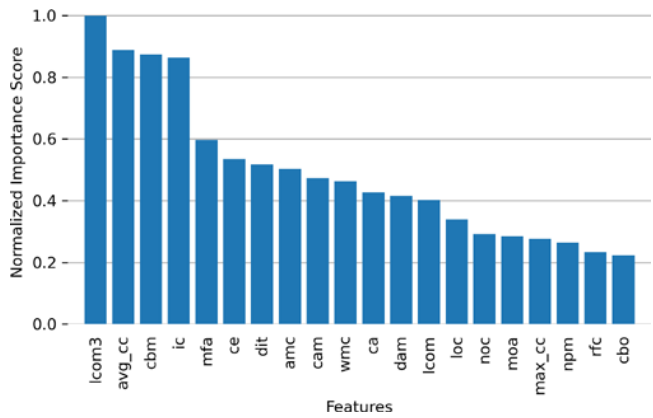

Fig. 2: Feature importance using the IG technique

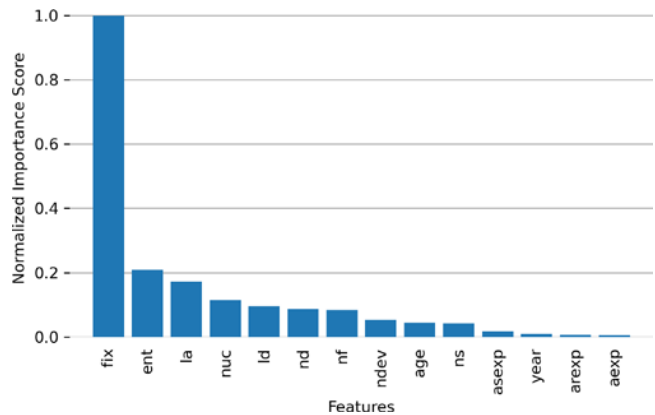

Fig. 4: Features importance from JIT model using IG technique.

Figure 4 shows the feature importance ranking from the Integrated Gradient technique on the JIT defect prediction model of the Camel project. This figure shows if the referred commit is a fix or not is the primary contributor in predicting whether this new commit can be bug-inducing or not. The next important feature is ent (distribution of modified code across each file) followed by la (number of lines added), nuc (number of unique changes), and nf (number of modified files). The next set of attributes is ld (lines of code deleted), ndev, age, and ns. It appears that the developer's experience (axep, sexp, and exp) did not have a significant contribution to the JIT model development from the IG approach. However, from Figure 5, we can notice that, unlike the IG method, the SHAP technique shows the total line added, and the developer experience used in the project strictly affects the performance of the SDP models.

Traditional and JIT SDP models showed different feature rankings when SHAP and IG methods were employed. The SHAP technique demonstrated that the lines of code used, and the total lines added in a program, are the most important feature for traditional and JIT SDP models respectively, whereas the IG technique highlighted the importance of other utilized features.

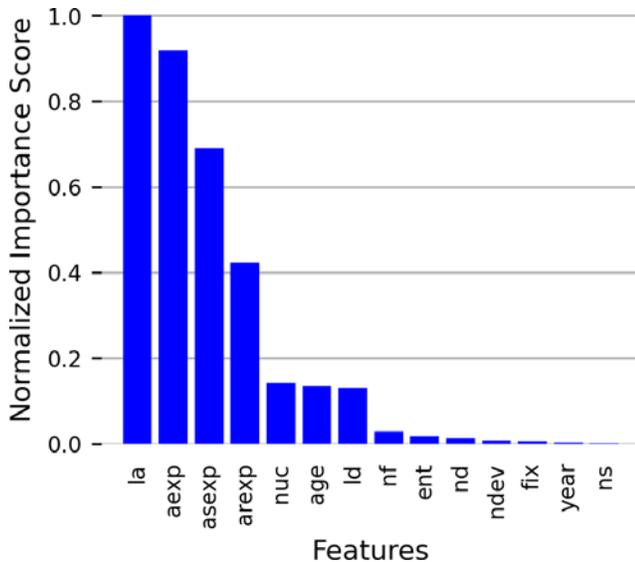

Fig. 5: Features importance from JIT model using SHAP.

This study can be extended by applying additional data-balancing strategies to see their impact on the performance and interoperability of the SDP models. In addition, data mining strategies can be applied for implementing additional features to evaluate their impacts on cross-project SDP models.

## V. CONCLUSION

This paper explored the development of traditional and Just-In-Time defect prediction models. For a balanced evaluation, the performance of the deep learning algorithm was compared with traditional machine learning algorithms. The SDP models were interpreted using model-specific Integrated Gradient and model-agnostic SHAP techniques. The feature importance rankings were compared and evaluated as they showed varied results. Applying this strategy to additional diverse projects can improve the generalization of these findings.

## VI. ACKNOWLEDGEMENT

We thank Ms. Mary Pierce, and Dr. Dev Sainani from Fanshawe College, and Dr. Kostas Kontogiannis for their support.